\documentclass[12pt]{article}
\title{Alternative evaluation of a $\ln \tan$ integral arising in quantum field theory}
\author{Mark W. Coffey\\
Department of Physics\\
Colorado School of Mines\\
Golden, CO  80401\\
(Received $\mbox{~~~~~~~~~~~~~~~~~~~~~~~~~~~~~~~2008}$)}
\date{November 9, 2008}
\begin{document}
\maketitle
\begin{center}

\baselineskip=25 pt
\begin{abstract}

A certain dilogarithmic integral $I_7$ turns up in a number of contexts including
Feynman diagram calculations, volumes of tetrahedra in hyperbolic geometry, knot 
theory, and conjectured relations in analytic number theory.  We provide an
alternative explicit evaluation of a parameterized family of integrals containing
this particular case.  By invoking the Bloch-Wigner form of the dilogarithm function,
we produce an equivalent result, giving a third evaluation of $I_7$.
We also alternatively formulate some conjectures which we
pose in terms of values of the specific Clausen function Cl$_2$.

\end{abstract}

\vfill
\baselineskip=15pt
\centerline{\bf Key words and phrases}
\medskip

\noindent
Clausen function, dilogarithm function, Hurwitz zeta function, functional equation,  duplication formula, triplication formula 
\end{center}

\bigskip
\centerline{\bf AMS classification numbers}
33B30, 11M35, 11M06   


\baselineskip=25pt
\pagebreak
\medskip

The particular integral
$$I_7 \equiv {{24} \over {7 \sqrt{7}}}\int_{\pi/3}^{\pi/2} \ln \left|{{\tan t+\sqrt{7}} \over {\tan t-\sqrt{7}}}\right |dt, \eqno(1)$$
occurs in a number of contexts and has received significant attention in the last several
years \cite{bb07,bbnotices,bbcomp,bb98}.
This and related integrals arise in hyperbolic geometry, knot theory, and quantum field
theory \cite{bb98,broad,broad2}.  
Very recently \cite{coffeyjmp} we obtained an explicit evaluation of (1) in terms of
the specific Clausen function Cl$_2$.  However, much work remains.  This is due to
the conjectured relation between a Dirichlet $L$ series and $I_7$ \cite{bb98},
$$I_7 \stackrel{?}{=}L_{-7}(2)=\sum_{n=0}^\infty\left [{1 \over {(7n+1)^2}}+
{1 \over {(7n+2)^2}}-{1 \over {(7n+3)^2}}+{1 \over {(7n+4)^2}}-{1 \over {(7n+5)^2}}-{1 \over {(7n+6)^2}}\right].  \eqno(2)$$
The ? here indicates that numerical verification to high precision has been performed but that no proof exists, the approximate numerical value of $I_7$ being
$I_7 \simeq 1.15192547054449104710169$.  
The statement (2) is equivalent to the conjecture, with
$\theta_7 \equiv 2\tan^{-1}\sqrt{7}$,
$${1\over 2}\left[3\mbox{Cl}_2(\theta_7)-3\mbox{Cl}_2(2\theta_7)+\mbox{Cl}_2(3\theta_7) \right]\stackrel{?}{=}{1 \over 4}Z_{Q(\sqrt{-7})}={7 \over 4}\left[\mbox{Cl}_2\left({{2\pi}
 \over 7}\right) +\mbox{Cl}_2\left({{4\pi} \over 7}\right)-\mbox{Cl}_2\left({{6\pi} \over 7}\right)\right],  \eqno(3)$$
relating triples of Clausen function values.
Here, we alternatively evaluate $I_7$ directly in terms of the left side of (3).
In addition, we present another evaluation of $I_7$, based upon a property
of the Bloch-Wigner form of the dilogarithm function.

We recall that the $L$ series $L_{-7}(s)$ has occurred in several places before,
including hyperbolic geometry \cite{zagier} and Dedekind sums of analytic number
theory \cite{almkvist}.  Let $\zeta_{Q(\sqrt{-p})}$ denote the Dedekind zeta function
of an imaginary quadratic field $Q(\sqrt{-p})$.  Then indeed we have 
\cite{almkvist,zagier,zucker}
$$\zeta_{Q(\sqrt{-7})}(s)={1 \over 2}\sum_{\stackrel{m,n \in Z}{(m,n) \neq (0,0)}}
{1 \over {(m^2+mn+2n^2)^s}} \eqno(4)$$
$$=\zeta(s)L_{-7}(s)=\zeta(s)7^{-s}\sum_{\nu=1}^6\left({\nu \over 7}\right)\zeta 
\left(s,{\nu \over 7}\right), \eqno(5)$$
where $\left({\nu \over 7}\right)$ is a Legendre symbol, $\zeta(s,a)$ is the Hurwitz
zeta function, and $\zeta(s)=\zeta(s,1)$ is the Riemann zeta function.

The series $L_{-7}(s)$ is an example of a Dirichlet $L$ function corresponding to a
real character $\chi_k$ [here, modulo 7] with $\chi_k(k-1)=-1$.
Such $L$ functions, extendable to the whole complex plane, satisfy the functional
equation \cite{zucker}
$$L_{-k}(s)={1 \over \pi}(2\pi)^s k^{-s+1/2}\cos\left({{s \pi} \over 2}\right)\Gamma
(1-s)L_{-k}(1-s). \eqno(6)$$
Owing to the relation $\Gamma(1-s)\Gamma(s)=\pi/\sin (\pi s)$,
this functional equation may also be written in the form
$$L_{-k}(1-s)=2(2\pi)^{-s} k^{s-1/2}\sin\left({{\pi s} \over 2}\right) \Gamma(s)
L_{-k}(s).  \eqno(7)$$
Integral representations are known for these $L$-functions \cite{zucker,coffeyjmp2}.  
From the functional equation (6) we find
$$\left.{\partial \over {\partial s}}L_{-k}(s)\right|_{s=-1}={k^{3/2} \over {4 \pi}}
L_{-k}(2).  \eqno(8)$$
In turn, we have 
$$\zeta_{Q(\sqrt{-k})}'(-1)=-{k^{3/2} \over {48\pi}}L_{-k}(2),  \eqno(9)$$
where we used $\zeta(-1)=-1/12$ and $L_{-k}(-1)=0$.

We have
{\newline \bf Proposition 1}.  
We have
$$I_7={4\over {7\sqrt{7}}}\left[3\mbox{Cl}_2(\theta_7)-3\mbox{Cl}_2(2\theta_7)
+\mbox{Cl}_2(3\theta_7)\right]. \eqno(10)$$

In fact, we treat integrals
$$I(a) \equiv \int_{\pi/3}^{\pi/2} \ln \left|{{\tan t+a} \over {\tan t-a}}\right |dt, \eqno(11)$$
and more general ones with varying limits.  For (11), we assume that $\pi/3 < \varphi
=\tan^{-1}a<\pi/2$.  These other integrals permit us to explicitly write other
conjectures directly in terms of linear combinations of specific Clausen function
values.  

The Clausen function Cl$_2$ can be defined by (e.g., \cite{lewin58,lewin1991})
$$\mbox{Cl}_2(\theta)\equiv -\int_0^\theta\ln\left|2 \sin {t \over 2}\right|dt
=\int_0^1 \tan^{-1}\left({{x \sin \theta} \over {1-x\cos \theta}}\right)
{{dx} \over x} \eqno(12)$$
$$=-\sin \theta \int_0^1 {{\ln x} \over {x^2-2x\cos \theta +1}}dx
=\sum_{n=1}^\infty {{\sin(n \theta)} \over n^2}.  \eqno(13)$$
When $\theta$ is a rational multiple of $\pi$ it is known that Cl$_2(\theta)$
may be written in terms of the trigamma and sine functions \cite{ded,grosjean}.
This Clausen function is odd and periodic, Cl$_2(\theta)=-\mbox{Cl}_2(-\theta)$,
and Cl$_2(\theta)=\mbox{Cl}_2(\theta+2\pi)$.  It also satisfies the duplication 
$${1 \over 2}\mbox{Cl}_2(2\theta)=\mbox{Cl}_2(\theta)-\mbox{Cl}_2(\pi-\theta),
\eqno(14)$$
triplication
$${1 \over 3}\mbox{Cl}_2(3\theta)=\mbox{Cl}_2(\theta)+\mbox{Cl}_2\left(\theta+{{2\pi}
\over 3}\right)+\mbox{Cl}_2\left(\theta+{{4\pi}\over 3}\right), \eqno(15)$$
and quadriplication 
$${1 \over 4}\mbox{Cl}_2(4\theta)=\mbox{Cl}_2(\theta)+\mbox{Cl}_2\left(\theta+{\pi
\over 2}\right)+\mbox{Cl}_2\left(\theta+\pi\right)+\mbox{Cl}_2\left(\theta+{{3\pi}
\over 2}\right), \eqno(16)$$
formulas, as well as a more general multiplication formula \cite{lewin58}. 
We recall the specific relation
$$\sum_{j=1}^6 \mbox{Cl}_2\left({{2 \pi} \over 7}j\right)=0, \eqno(17)$$
that arises as a special case of \cite{lewin58} (pp. 95, 253)
$$\sum_{j=1}^{n-1} \mbox{Cl}_2\left({{2 \pi} \over n}j\right)=0. \eqno(18)$$
In (17), pairwise cancellation occurs, as Cl$_2(\theta)=-\mbox{Cl}_2(2\pi-\theta)$.

Further information on the special functions that we employ may readily be found
elsewhere \cite{lewin,lewin1991,sri,coffeyjmp2}.  In particular, with 
$$\mbox{Li}_2(z)=\sum_{k=1}^\infty {z^k \over k^2}, ~~~~|z| \leq 1, \eqno(19)$$
or
$$\mbox{Li}_2(z)=-\int_0^z {{\ln(1-t)} \over t}dt, \eqno(20)$$
the dilogarithm function, we have the relation
$$\mbox{Li}_2(e^{i\theta})={\pi^2 \over 6}-{1 \over 4}\theta(2\pi-\theta)+i\mbox{Cl}_2
(\theta), ~~~~~~0 \leq \theta \leq 2\pi.  \eqno(21)$$
We omit discussion of further relations between the Clausen function Cl$_2$ and
the dilogarithm function.  


For the proof of Proposition 1 we repeatedly rely on \cite{lewin58} (pp. 227, 272)
$$\int_0^\theta \ln(\tan \theta+\tan \varphi)d\theta=-\theta\ln(\cos \varphi)-{1 \over 2}
\mbox{Cl}_2(2\theta+2\varphi)+{1 \over 2}\mbox{Cl}_2(2\varphi)-{1 \over 2}\mbox{Cl}_2
(\pi-2\theta).  \eqno(22)$$
We may split the integral in (11), writing
$$I(a) = \int_{\pi/3}^\varphi \ln \left({{a+\tan t} \over {a-\tan t}}\right)dt+
\int_\varphi^{\pi/2} \ln \left({{\tan t+a} \over {\tan t-a}}\right)dt$$
$$=\int_{\pi/3}^{\pi/2} \ln(a+\tan t)dt-\int_{\pi/3}^\varphi \ln(a-\tan t)dt
-\int_\varphi^{\pi/2}\ln(\tan t-a)dt.  \eqno(23)$$
By the use of (22) we obtain for the first integral on the right side of (23)
$$\int_{\pi/3}^{\pi/2} \ln(a+\tan t)dt=-{\pi \over 6}\ln \cos \varphi +{1 \over 2}\left[\mbox{Cl}_2\left({{2 \pi} \over 3}
+2\varphi\right)-\mbox{Cl}_2\left(\pi+2\varphi\right)\right]+{1 \over 2}\mbox{Cl}_2
\left({\pi \over 3}\right),  \eqno(24a)$$
the second integral,
$$\int_{\pi/3}^\varphi \ln(a-\tan t)dt=-\left(\varphi-{\pi \over 3}\right)\ln \cos\varphi
-{1 \over 2}\mbox{Cl}_2\left[2\left(\varphi-{\pi \over 3}\right)\right]+{1 \over 2}\left[\mbox{Cl}_2\left({\pi \over 3}\right)-\mbox{Cl}_2(\pi-2\varphi)\right], \eqno(24b)$$
and the third integral,
$$\int_\varphi^{\pi/2}\ln(\tan t-a)dt=-\left({\pi \over 2}-\varphi\right)\ln \cos\varphi
=-\cot^{-1} a \ln \cos \varphi.  \eqno(24c)$$
This latter integral is readily obtained from (22) by taking $a \to -a$ so that 
simply $\tan \varphi \to -\tan \varphi$.
Then per (23) we have
$$I(a)={1 \over 2}\left[\mbox{Cl}_2\left(2\varphi+{{2\pi} \over 3}\right)+ \mbox{Cl}_2 \left(2\varphi-{{2\pi} \over 3}\right)\right]-\mbox{Cl}_2(\pi+2\varphi).  \eqno(25)$$
Then we apply both the duplication formula (14) and the triplication formula (15)
wherein Cl$_2(\theta+4\pi/3)=\mbox{Cl}_2(\theta-2\pi/3)$ as Cl$_2(\theta)=\mbox{Cl}_2
(\theta-2\pi)$ by the $2\pi$-periodicity of Cl$_2$.  We find
$$I(a)={1 \over 6}\left[\mbox{Cl}_2(6\varphi)-3\mbox{Cl}_2(4\varphi)+3\mbox{Cl}_2
(2\varphi)\right].  \eqno(26)$$
When $\varphi=\tan^{-1} \sqrt{7}$, the case (10) follows.

We next present some reference integrals.  We then apply them to write expressions for
combinations of the integrals
$$I_n \equiv \int_{n\pi/24}^{(n+1)\pi/24} \ln \left|{{\tan t+\sqrt{7}} \over {\tan t-\sqrt{7}}}\right |dt, \eqno(27)$$
where $n \geq 0$ is an integer.  We supplement (22) with
$$\int_x^y \ln(a-\tan t)dt=-(y-x)\ln \cos\varphi+{1 \over 2}\left[\mbox{Cl}_2[2(\varphi-y)]
-\mbox{Cl}_2[2(\varphi-x)]\right]$$
$$-{1 \over 2}\left[\mbox{Cl}_2(\pi -2y)-\mbox{Cl}_2(\pi-2x)\right], \eqno(28)$$
where $a=\tan \varphi$.  We also have 
$$\int_x^y \ln \left|{{\tan t+a} \over {\tan t-a}}\right|dt={1 \over 2} \left[\mbox{Cl}_2(2x+2\varphi)-\mbox{Cl}_2(2x-2\varphi)+\mbox{Cl}_2(2y-2\varphi)
-\mbox{Cl}_2(2y+2\varphi)\right], \eqno(29)$$
with $x < \varphi =\tan^{-1} a < y$.
We now write expressions for the linear combinations
$$C_1 \equiv -2(I_2+I_3+I_4+I_5)+I_8+I_9-(I_{10}+I_{11}) \stackrel{?}{=}0, \eqno(30)$$
and
$$C_2 \equiv I_2+3(I_3+I_4+I_5)+2(I_6+I_7)-3I_8-I_9 \stackrel{?}{=}0. \eqno(31)$$
These relations have been detected with further PSLQ computations \cite{bbcomp}.
A similar conjecture for integrals $I_n$ with increments $n\pi/60$ has also been
written \cite{bbnotices} (p. 508).  The latter linear combination may also be expressed
in terms of Cl$_2$ values, but here we concentrate on (30) and (31). We decompose the
left side of (30) as indicated, and find for these contributions
$$2(I_2+I_3+I_4+I_5)=\mbox{Cl}_2\left({\pi \over 6}+2\varphi\right)-
\mbox{Cl}_2\left({\pi \over 6}-2\varphi\right)+\mbox{Cl}_2\left({\pi \over 2}-2\varphi\right)-\mbox{Cl}_2\left({\pi \over 2}+2\varphi\right), \eqno(32)$$
where $\varphi=\tan^{-1} \sqrt{7}$,
$$2I_8=\mbox{Cl}_2\left({{2\pi} \over 3}+2\varphi\right)-
\mbox{Cl}_2\left({{3\pi} \over 4}+2\varphi\right)+\mbox{Cl}_2\left(2\varphi-{{2\pi} \over 3}\right)
-\mbox{Cl}_2\left(2\varphi-{{3\pi} \over 4}\right), \eqno(33)$$
$$2I_9=\mbox{Cl}_2\left({{3\pi} \over 4}+2\varphi\right)-
\mbox{Cl}_2\left({{3\pi} \over 4}-2\varphi\right)+\mbox{Cl}_2\left({{5\pi} \over 6}-2\varphi\right)-\mbox{Cl}_2\left({{5\pi} \over 6}+2\varphi\right), \eqno(34)$$
and
$$I_{10}+I_{11}=-\mbox{Cl}_2(\pi+2\varphi)+{1 \over 2}\left[\mbox{Cl}_2\left({{5\pi} 
\over 6}+2\varphi\right)-\mbox{Cl}_2\left({{5\pi} \over 6}-2\varphi\right)\right]. \eqno(35)$$
Therefore, we obtain 
$$C_1=-\mbox{Cl}_2\left({\pi \over 6}+2\varphi\right)+
\mbox{Cl}_2\left({\pi \over 6}-2\varphi\right)-\mbox{Cl}_2\left({\pi \over 2}-2\varphi\right)+\mbox{Cl}_2\left({\pi \over 2}+2\varphi\right)$$
$${1 \over 2}\left[\mbox{Cl}_2\left({{2\pi} \over 3}+2\varphi\right)
+\mbox{Cl}_2\left(2\varphi-{{2\pi} \over3}\right)\right]
+\mbox{Cl}_2\left({{5\pi} \over 6}-2\varphi\right)-\mbox{Cl}_2\left({{5\pi} \over 6}+2\varphi\right)$$
$$+\mbox{Cl}_2(\pi+2\varphi).  \eqno(36)$$

For the combination $C_2$ we have
$$2I_2=\mbox{Cl}_2\left({\pi \over 6}+2\varphi\right)-
\mbox{Cl}_2\left({\pi \over 6}-2\varphi\right)+\mbox{Cl}_2\left({\pi \over 4}-2\varphi\right)-\mbox{Cl}_2\left({\pi \over 4}+2\varphi\right), \eqno(37)$$
$$-2(I_6+I_7)=\mbox{Cl}_2\left({{2\pi} \over 3}+2\varphi\right)-
\mbox{Cl}_2\left({{2\pi} \over 3}-2\varphi\right)+\mbox{Cl}_2\left({\pi \over 2}-2\varphi\right)-\mbox{Cl}_2\left({\pi \over 2}+2\varphi\right), \eqno(38)$$
and
$$2(I_3+I_4+I_5)=\mbox{Cl}_2\left({\pi \over 4}+2\varphi\right)-
\mbox{Cl}_2\left({\pi \over 4}-2\varphi\right)+\mbox{Cl}_2\left({\pi \over 2}-2\varphi\right)-\mbox{Cl}_2\left({\pi \over 2}+2\varphi\right). \eqno(39)$$
Therefore, we find 
$$2C_2=\mbox{Cl}_2\left({\pi \over 6}+2\varphi\right)-
\mbox{Cl}_2\left({\pi \over 6}-2\varphi\right)+\mbox{Cl}_2\left({\pi \over 2}-2\varphi\right)-\mbox{Cl}_2\left({\pi \over 2}+2\varphi\right)$$
$$-5\mbox{Cl}_2\left({{2\pi} \over 3}+2\varphi\right)
-5\mbox{Cl}_2\left(2\varphi-{{2\pi} \over 3}\right)
-\mbox{Cl}_2\left({{5\pi} \over 6}-2\varphi\right)+\mbox{Cl}_2\left({{5\pi} \over 6}+2\varphi\right)$$
$$+2\left[\mbox{Cl}_2\left({\pi \over 4}+2\varphi\right)-
\mbox{Cl}_2\left({\pi \over 4}-2\varphi\right)+\mbox{Cl}_2\left({{3\pi} \over 4}+2\varphi\right)+\mbox{Cl}_2\left(2\varphi-{{3\pi} \over 4}\right)\right].
\eqno(40)$$

By a combination of the quadriplication formula (16) and the duplication formula (14)
we may write
$${1 \over 4}\mbox{Cl}_2(4\theta)=\mbox{Cl}_2\left(\theta+{\pi \over 2}\right)+
\mbox{Cl}_2\left(\theta-{\pi \over 2}\right)+{1 \over 2}\mbox{Cl}_2(2\theta).
\eqno(41)$$
This enables other expressions for $C_1$ and $C_2$.  Similarly, one may use the 
$6$- and $12$-fold multiplication formulas.

In regard to the combination on the right side of (3), we comment on an observation
given previously \cite{coffeyjmp}.  We have that $\pm \sin (2\pi/7)$, $\pm \sin (4\pi/7)$, 
and $\pm \sin (6\pi/7)$ are the nonzero roots of the Chebyshev polynomial
$T_7(x)$.  Indeed, if we write the cubic polynomials
$$p_1(x)=\left(x- \sin {{2 \pi} \over 7}\right)\left(x- \sin {{4 \pi} \over 7}\right)\left(x+ \sin {{6 \pi} \over 7}\right)=x^3-{\sqrt{7} \over 2}x^2+{\sqrt{7}
\over 8}, \eqno(42a)$$
and
$$p_2(x)=\left(x- \sin {{6 \pi} \over 7}\right)\left(x+ \sin {{2 \pi} \over 7}\right)\left(x+ \sin {{4 \pi} \over 7}\right)=x^3+{\sqrt{7} \over 2}x^2-{\sqrt{7}
\over 8}, \eqno(42b)$$
we then have the factorization $p_1(x)p_2(x)=T_7(x)/64x$.  This invites questions
as to whether scaled versions of these or other Chebyshev polynomials could be
useful in developing identities underlying (3), (30), (31), or the like.

Given the close relation of the Clausen function Cl$_2$ and the dilogarithm function,
one wonders if a set of ladder relations for the latter may be carried over to explain
(3) and relations amongst the integrals $I_n$.  In developing ladder relations,
cyclotomic equations for the base have proven very useful.  It would be of interest
to see if Cl$_2$ relations with $\theta_7$ could be discovered in this way.

We remark on using Kummer's relation \cite{lewin58} (pp. 107, 254) to rewrite
the right side of (3) in terms of the dilogarithm of complex argument.  We have
$${1 \over 4}Z_{Q(\sqrt{-7})}={7 \over 2}\left[\mbox{Im}~ \mbox{Li}_2(Re^{i\phi})-
b\ln R\right], \eqno(43)$$
where
$$R={{\tan b} \over {\sin \phi +\tan b \cos \phi}}.  \eqno(44)$$
Here, we may take $\phi=\pi/7$ and $b=2\pi/7$, or vice versa.  Then by Proposition 2
of \cite{coffeyjmp} we have the integral representation
$${\sqrt{7} \over 2}I_7\stackrel{?}{=}\mbox{Cl}_2\left({{2\pi}
 \over 7}\right) +\mbox{Cl}_2\left({{4\pi} \over 7}\right)-\mbox{Cl}_2\left({{6\pi} \over 7}\right)=2\sin\left({\pi \over 7}\right)\int_d^\infty {{\ln y ~ dy} \over {y^2-2y
\cos(\pi/7)+1}}, \eqno(45)$$
where $d=[2\cos(\pi/7)-1]^{-1}$.

Finally, we use relations from \cite{lewin1991} (Appendix A) and \cite{zagier} to
write a third evaluation of the integral $I_7$.  For this we introduce the angle
$\theta_{75}\equiv 2\tan^{-1}(\sqrt{7}/5)$ and the Bloch-Wigner dilogarithm \cite{ftnote}
$$D(z)=\mbox{Im}[\mbox{Li}_2(z)]+\mbox{arg}(1-z)\ln |z|, \eqno(46)$$
for which we have \cite{lewin1991} (p. 246)
$$D(z)={1 \over 2}[\mbox{Cl}_2(2\theta)+\mbox{Cl}_2(2\omega)-\mbox{Cl}_2(2\theta+
2\omega)],  \eqno(47)$$
where $\theta=\mbox{arg} ~z$ and $\omega=\mbox{arg} ~(1-\bar{z})$.  We note the
interpretation that for $z \in C$, the volume of the asymptotic simplex with vertices $0, 1, z$, and $\infty$ in $3$-dimensional hyperbolic space is given by $|D(z)|$ \cite{lewin1991} (p. 271).  We then rewrite
the expression (\cite{lewin1991}, p. 384 or \cite{zagier}, p. 246)
$$\zeta_{Q(\sqrt{-7})}(2)=\zeta(2)L_{-7}(2)={{4\pi^2} \over {21\sqrt{7}}}\left[2D\left(
{{1+i\sqrt{7}} \over 2}\right)+D\left({{-1+i\sqrt{7}} \over 4}\right)\right].  \eqno(48)$$
We apply (47), giving
$$I_7\stackrel{?}{=}L_{-7}(2)={8 \over {7\sqrt{7}}}\left[2D\left(
{{1+i\sqrt{7}} \over 2}\right)+D\left({{-1+i\sqrt{7}} \over 4}\right)\right]$$
$$={4 \over {7\sqrt{7}}}\left[4\mbox{Cl}_2(\pi-\theta_7)-\mbox{Cl}_2(\theta_7)+
\mbox{Cl}_2(\theta_{75})+\mbox{Cl}_2(\theta_7-\theta_{75})\right].  \eqno(49)$$
In the case of $D[(1+i\sqrt{7})/2]$ we used the duplication formula (14).
In contrast to (49), the expression in \cite{coffeyjmp} for $I_7$ involves
$\theta_+ \equiv \tan^{-1} (\sqrt{7}/3)$.  
With the various analytic evaluations now known for $I_7$ or $L_{-7}(2)$, we have 
enlarged the set of possible relations amongst Cl$_2$ values.  From (10), (14), and
(49) we obtain the conjecture
$$\mbox{Cl}_2(3\theta_7)-\mbox{Cl}_2(2\theta_7)\stackrel{?}{=}
\mbox{Cl}_2(\theta_{75})+\mbox{Cl}_2(\theta_7-\theta_{75}).  \eqno(50)$$
In fact, we have $\theta_7-\theta_{75}=2\theta_+$, and we conclude by proving (50),
and thereby (49).  We quickly show that both
$$\mbox{Cl}_2(3\theta_7)=\mbox{Cl}_2(\theta_{75}) \eqno(51)$$
and
$$\mbox{Cl}_2(2\theta_7)=-\mbox{Cl}_2(\theta_7-\theta_{75}),  \eqno(52)$$
for we have $3\theta_7-2\pi=\theta_{75}$ and $\theta_7-\pi=-\theta_+$.  The latter
relations require nothing more than the identity $\tan(x/2)=\sin x/(1+\cos x)$.

We have similarly found many other angular pairs $(\theta_1,\theta_2)$ 
satisfying $3\theta_1-2\pi=\pm \theta_2$, immediately giving Cl$_2(3\theta_1)=
\pm \mbox{Cl}_2(\theta_2)$.  As these may be useful elsewhere \cite{bb98,lewin1991}, 
we record several of them in the first Appendix.  We also relegate to this Appendix a
possibly new log trigonometric integral in terms of Cl$_2$.  In the second Appendix,
we develop new series and integral representations of the Clausen function.

\medskip
\centerline{\bf Appendix A}
\medskip

We let $\theta_k \equiv 2\tan^{-1} \sqrt{k}$, and $\theta_{k,j} \equiv 2\tan^{-1}\sqrt{k}/j$. 
We find the relations
$$3\theta_2-2\pi=-\theta_{2,5}, ~~~~~~3\theta_5-2\pi=\theta_{5,7}, \eqno(A.1)$$
$$3\theta_{11}-2\pi=-\theta_{11,4}, ~~~~~~3\theta_{13}-2\pi=-2\tan^{-1}\left({{5\sqrt{13}}
\over {19}}\right), \eqno(A.2)$$
and
$$3\theta_{91,3}-2\pi=-2\tan^{-1}\left({{8\sqrt{91}}\over {99}}\right), ~~~~~~
3\theta_{91,5}-2\pi=2\tan^{-1}\left({{2\sqrt{91}}\over {155}}\right), ~~~~~~
3\theta_{91,7}-2\pi=-\theta_{91,28}. \eqno(A.3)$$
With $\theta_{32}\equiv 2\tan^{-1}\sqrt{3/2}$, $\theta_{53}\equiv 2\tan^{-1}\sqrt{5/3}$,
$\theta_{133}\equiv 2\tan^{-1}\sqrt{13/3}$, $\theta_{73}\equiv 2\tan^{-1}\sqrt{7/3}$,
we have
$$3\theta_{32}-2\pi=-2\tan^{-1}\left({3 \over 7}\sqrt{3 \over 2}\right), ~~~~~~
3\theta_{53}-2\pi=-2\tan^{-1}\left({1 \over 3}\sqrt{5 \over 3}\right),  \eqno(A.4)$$
and
$$3\theta_{133}-2\pi=2\tan^{-1}\left({1 \over 9}\sqrt{{13} \over 3}\right), ~~~~~~
3\theta_{73}-2\pi=-2\tan^{-1}\left({1 \over 9}\sqrt{7 \over 3}\right).  \eqno(A.5)$$

Based upon the trigonometric identity $3+4\cos\theta+\cos2\theta=2(1+\cos\theta)^2$,
we have found the integral
$$\int_0^x \ln(3+4\cos\theta+\cos2\theta)d\theta=-x\ln 2+4\mbox{Cl}_2(\pi-x), ~~~~
0 \leq x \leq \pi.  \eqno(A.6)$$
This obviously provides an integral expression for the Catalan constant
$G=$ \newline{$\sum_{k\geq 0}(-1)^k/(2k+1)^2=\mbox{Cl}_2(\pi/2)$ when $x=\pi/2$.}

The function Cl$_2(t)$ for $t\in(0,\pi)$ has its only maximum at $t=\pi/3$, when
Cl$_2(\pi/3) \simeq 1.014941606409653625021$.  We mention that near this value Cl$_2$
has a fixed point, Cl$_2(y)=y$ for $y \simeq 1.01447193895251725798414$.

Related to the equality of expressions (10) and (49) for $I_7$ we have the relation
$$6\left[\mbox{Li}_2\left({{1-3i\sqrt{7}} \over 8}\right)+\mbox{Li}_2\left({{1+3i\sqrt{7}} \over 8}\right)\right]=3(\pi-2\theta_+)^2-\pi^2$$
$$=3(\theta_7-\theta_+)^2-\pi^2=3[\pi-\tan^{-1}(3\sqrt{7})]^2-\pi^2.  \eqno(A.7)$$
Such relations follow readily from (21) as we have
$$6[\mbox{Li}_2(e^{i\theta})+\mbox{Li}_2(e^{-i\theta})]=2\pi^2+3\theta^2,
~~~~~~ 0 \leq \theta \leq 2\pi.  \eqno(A.8)$$

\pagebreak
\centerline{\bf Appendix B}
\medskip

We have
{\newline \bf Proposition B1}.  We have for $\theta<\pi/n$ and $n \geq 1$ an integer
$${1 \over 2}\left[{1 \over n}\mbox{Cl}_2(2n\theta)-\mbox{Cl}_2(2\theta)\right]
=\sum_{j=1}^\infty {{\zeta(2j)} \over {j\pi^{2j}}} {{(n^{2j}-1)} \over {(2j+1)}}
\theta^{2j+1}-\theta \ln n. \eqno(B.1)$$
This result gives several Corollaries, including
{\newline Corollary (i)}
$${1 \over 2}\sum_{k=1}^{n-1} \mbox{Cl}_2\left(2\theta+{{2\pi} \over n}k\right)
=\sum_{j=1}^\infty {{\zeta(2j)} \over {j\pi^{2j}}} {{(n^{2j}-1)} \over {(2j+1)}}
\theta^{2j+1}-\theta \ln n, \eqno(B.2)$$
{\newline Corollary (ii)}
$${1 \over 2}\left[{1 \over n}\mbox{Cl}_2(2n\theta)-\mbox{Cl}_2(2\theta)\right]
=-\theta \ln n$$
$$+{{2\theta} \over n}\int_0^\infty {1 \over x^2}[\sinh(nx)-n \sinh x]{{dx} \over
{(e^{\pi x/\theta}-1)}}, \eqno(B.3)$$
{\newline Corollary (iii)}
$${1 \over 2}\left[{1 \over n}\mbox{Cl}_2(2n\theta)-\mbox{Cl}_2(2\theta)\right]
=-\theta \ln n$$
$$+{\theta \over {2\pi}}\left[2n\theta \tanh^{-1}\left({{n\theta} \over \pi}\right)
-2\theta  \tanh^{-1}\left({\theta \over \pi}\right)+\pi \ln\left({{\pi^2-n^2\theta^2}
\over {\pi^2-\theta^2}}\right)\right]$$
$$+2\pi\int_1^\infty \left[\tanh^{-1}\left({\theta \over {\pi x}}\right)-{1 \over n}
\tanh^{-1}\left({{n\theta} \over {\pi x}}\right)\right]P_1(x)dx.  \eqno(B.4)$$
In the last equation, $P_1(x)=x-[x]-1/2$ is the first periodized Bernoulli polynomial.

{\it Proof}.  The Proposition is based upon the relation \cite{nbs} (p. 75)
$$\ln\left({{n \sin x} \over {\sin nx}}\right)=\sum_{j=1}^\infty {{\zeta(2j)} \over
{j \pi^{2j}}} (n^{2j}-1)x^{2j}, ~~~~~~ |x| < \pi/n,  \eqno(B.5)$$
wherein we have used the relation between $\zeta(2j)$ and the Bernoulli numbers
$B_{2j}$.  (For more details, see the end of this Appendix.)
With the series of (B.5) being boundedly convergent, we may integrate 
term-by-term over any finite interval avoiding the singularity at $x=\pi/n$.
Doing so, integrating over $[0,\theta]$, and using the first integral representation
for Cl$_2$ on the right side of (12) gives (B.1).

Corollary (i) follows from the multiplication formula for Cl$_2$ \cite{lewin58}
(pp. 94, 253).  Corollary (ii) uses a standard integral representation of the 
Riemann zeta function.  With the interchange of summation and integration, with
the integral being absolutely convergent, the Corollary follows.

Corollary (iii) uses the representation for Re $s>-1$,
$$\zeta(s)={1 \over 2}+{1 \over {s-1}}-s\int_1^\infty {{P_1(x)} \over x^{s+1}}dx.
\eqno(B.6)$$
Again, the interchange of summation and integration is employed.

{\it Remarks}.  In connection with (B.5) we may note the relation with the
Chebyshev polynomials $U_n$ of the second kind \cite{grad} (p. 1032),
$$U_{n-1}(\cos \phi)= {{\sin n\phi} \over {\sin \phi}}.  \eqno(B.7)$$

For $\theta=\pi/4$ and other values, Proposition B1 gives many relations involving
the Catalan constant $G$.  More generally, for $\theta$ a rational multiple of
$\pi$, the results are expressible in terms of $\psi'$, the trigamma function
\cite{ded,grosjean}.  If we let
$$r(\theta,n) \equiv \sum_{j=1}^\infty {{\zeta(2j)} \over {j\pi^{2j}}} {{(n^{2j}-1)} \over {(2j+1)}} \theta^{2j+1}, ~~~~~~ \theta \leq \pi/n, \eqno(B.8)$$
we may write several simple examples:
$$r\left({\pi \over 3},2\right)=\pi\left[{1 \over {18}}(\sqrt{3}\pi+6\ln 2)-{{\psi'(1/3)}
\over {4\sqrt{3}\pi}}\right], \eqno(B.9a)$$
$$r\left({\pi \over 3},3\right)=\pi\left[{1 \over {27}}(\sqrt{3}\pi+9\ln 3)-{{\psi'(1/3)}
\over {6\sqrt{3}\pi}}\right], \eqno(B.9b)$$
$$r\left({\pi \over 4},2\right)=-{1 \over 2}G+{\pi \over 4}\ln 2, \eqno(B.10a)$$
$$r\left({\pi \over 4},3\right)=-{2 \over 3}G+{\pi \over 4}\ln 3, \eqno(B.10b)$$
$$r\left({\pi \over 4},4\right)=-{1 \over 2}G+{\pi \over 2}\ln 2, \eqno(B.10c)$$
and
$$r\left({\pi \over 6},2\right)=\pi\left[{1 \over {54}}(2\sqrt{3}\pi+9\ln 2)-{{\psi'(1/3)}
\over {6\sqrt{3}\pi}}\right], \eqno(B.11a)$$
$$r\left({\pi \over 6},3\right)=\pi\left[{1 \over {18}}(\sqrt{3}\pi+3\ln 3)-{{\psi'(1/3)}
\over {4\sqrt{3}\pi}}\right], \eqno(B.11b)$$
$$r\left({\pi \over 6},4\right)=\pi\left[{1 \over {108}}(7\sqrt{3}\pi+36\ln 2)-{{7\psi'(1/3)}\over {24\sqrt{3}\pi}}\right], \eqno(B.11c)$$
$$r\left({\pi \over 6},5\right)=\pi\left[{1 \over {30}}(2\sqrt{3}\pi+5\ln 5)-{{\sqrt{3}\psi'(1/3)} \over {10\pi}}\right], \eqno(B.11d)$$
$$r\left({\pi \over 6},6\right)=\pi\left[{1 \over {18}}(\sqrt{3}\pi+3\ln 2+3\ln 3) -{{\psi'(1/3)}\over {4\sqrt{3}\pi}}\right]. \eqno(B.11e)$$

Finally, we supply a derivation of (B.5).  We have
$${d \over {dx}} \ln\left({{n \sin x} \over {\sin nx}}\right)=\cot x - n \cot nx$$
$$=\sum_{k=1}^\infty {{2^{2k} |B_{2k}|} \over {(2k)!}}(n^{2k}-1)x^{2k-1}, ~~~~ {{n|x|} \over \pi} < 1, \eqno(B.12)$$
where we used a series representation for $\cot$ \cite{grad} (p. 35).  Since
$${{2^{2k} |B_{2k}|} \over {(2k)!}}={{2\zeta(2k)} \over \pi^{2k}}, \eqno(B.13)$$
we have
$$\cot x - n \cot nx=2\sum_{k=1}^\infty {{\zeta(2k)} \over \pi^{2k}}(n^{2k}-1)x^{2k-1}.
\eqno(B.14)$$
Integrating both sides of this relation gives (B.5).


\pagebreak

\end{document}